\documentclass[a4paper,fleqn]{cas-dc}

\usepackage[numbers]{natbib}

\def\tsc#1{\csdef{#1}{\textsc{\lowercase{#1}}\xspace}}
\tsc{WGM}
\tsc{QE}
\tsc{EP}
\tsc{PMS}
\tsc{BEC}
\tsc{DE}

\begin{document}
\let\WriteBookmarks\relax
\def\floatpagepagefraction{1}
\def\textpagefraction{.001}

\shorttitle{}    

\shortauthors{B. K. Sirasva et~al }  

\title [mode = title]{A cosmic muon tomography system with machine learning based momentum measurement for multi-object reconstruction and material characterization}  



%

\author[1]{Bharat Kumar Sirasva}[
orcid=0009-0009-8508-9641
]
\cormark[1]
\ead{bharatsirswa111@gmail.com}
\credit{Conceptualization, Methodology, Software, Investigation, Data curation, Formal analysis, Visualization, Writing -- original draft}
\author[2]{Rohit Gupta}[
orcid=0000-0002-8398-680X
]
\credit{Validation, Investigation, Writing -- review and editing}
\author[1]{Satyajit Jena}[
orcid=0000-0002-6220-6982
]
\cormark[2]
\ead{sjena@iisermohali.ac.in}
\credit{Supervision, Writing -- review}
\affiliation[1]{
    organization={Department of Physical Sciences, Indian Institute of Science Education and Research Mohali},
    city={Mohali},
    postcode={140306},
    state={Punjab},
    country={India}
}

\affiliation[2]{
    organization={Department of Physics, Faculty of Science, Kashi Naresh University},
    city={Bhadohi},
    postcode={221304},
    state={Uttar Pradesh},
    country={India}
}


\cortext[cor1]{Corresponding author}
\cortext[cor2]{Principal corresponding author}

\begin{abstract}
Cosmic muon tomography is a powerful non-destructive imaging technique for inspecting dense and shielded materials through multiple Coulomb scattering. In this work, we present the design, simulation, and performance evaluation of a complete muon tomography system comprising six scintillator-strip tracking stations for trajectory reconstruction and a four-station magnetic spectrometer for muon momentum estimation. The detector geometry is implemented in the GEANT4 framework and optimized for object localization and material characterization. The reconstructed momentum is combined with the scattering angle to define the scattering density $\rho_s = {(\theta p)^2}/{L_{\mathrm{eff}}}$, which enhances sensitivity to material-dependent scattering. Point-of-Closest-Approach (PoCA) reconstruction is used to estimate scattering locations within the imaging volume. To detect and separate multiple unknown objects, Hierarchical Density-Based Spatial Clustering of Applications with Noise (HDBSCAN) is applied to the reconstructed PoCA cloud. Cluster-level scattering and geometric features are then extracted for object characterization. The proposed framework enables object detection, localization, volume estimation, shape reconstruction, and material ranking within a unified analysis pipeline. Simulation studies with multiple objects of different compositions demonstrate accurate reconstruction of object positions and geometries, while providing reliable material discrimination based on scattering density. The developed system offers a scalable approach for next-generation cosmic muon tomography applications in security screening, nuclear waste characterization, and non-destructive inspection.
\end{abstract}




\maketitle

\section{Introduction}

The ability to investigate the internal structure of large, dense, or shielded objects without physical intrusion is of considerable importance in a wide range of scientific, industrial, and security-related applications. These include cargo inspection, nuclear waste characterization, geological exploration, archaeological studies, and non-destructive testing of critical infrastructure \cite{schultz2007,Bonechi:2019ckl,HOU2026104091,PRXEnergy.4.013002}. Conventional imaging techniques such as $X$-ray radiography and computed tomography often encounter significant limitations when probing thick or high-density materials due to their limited penetration capability and the requirement for artificial radiation sources. In contrast, cosmic muon tomography has emerged as a powerful passive imaging technique that exploits naturally occurring atmospheric muons as highly penetrating probes of matter ~\cite{borozdin2003,Morishima:2017ghw,Schultz:2004kx,MORRIS15102008}.

Cosmic-ray muons are continuously generated in the Earth's atmosphere through interactions between primary cosmic rays and atmospheric nuclei \cite{Gaisser:2016uoy}. These interactions produce secondary mesons, primarily charged pions and kaons, which subsequently decay into muons \cite{ParticleDataGroup:2024cfk,Gaisser:2016uoy}. Owing to their relatively long lifetime and relativistic velocities, a substantial fraction of these muons reach the Earth's surface before decaying. The resulting muon flux at sea level is approximately one particle per square centimeter per minute, providing a naturally available source of penetrating radiation without the need for dedicated particle accelerators or radioactive sources \cite{ParticleDataGroup:2014cgo}. Because muons can traverse several meters of dense material, they are particularly well suited for imaging applications involving large or heavily shielded objects \cite{borozdin2003,Morishima:2017ghw,MORRIS15102008,schultz2007}.

The physical principle underlying muon tomography is based on the phenomenon of multiple Coulomb scattering. As a muon traverses matter, it undergoes numerous small-angle electromagnetic interactions with atomic nuclei, resulting in a net angular deflection from its original trajectory \cite{Highland:1975pq,LYNCH19916,ParticleDataGroup:2024cfk}. The magnitude of this scattering depends on both the properties of the traversed material and the momentum of the incident muon. The root-mean-square scattering angle can be approximated using the Highland formula

\begin{equation}
\theta_{0}
=
\frac{13.6~\mathrm{MeV}}
{\beta p c}
\sqrt{\frac{x}{X_0}}
\left[
1+0.038
\ln
\left(
\frac{x}{X_0}
\right)
\right],
\end{equation}

where $p$ is the muon momentum, $\beta=v/c$, $x$ is the material thickness, and $X_0$ is the radiation length of the material. Since materials with high atomic number generally possess smaller radiation lengths, they induce larger scattering angles than low-$Z$ materials. Consequently, measurements of muon deflections can be used to infer both the location and composition of materials within an unknown volume.

Muon tomography systems reconstruct the trajectories of cosmic-ray muons before and after they traverse an interrogation volume, allowing the scattering location and scattering strength to be estimated from the measured track deflections. Several reconstruction approaches have been proposed in the literature, including maximum-likelihood reconstruction \cite{schultz2007}, statistical inversion methods \cite{10.1063/5.0273072}, and voxel-based imaging algorithms \cite{Schultz:2004kx}. Among these techniques, the Point-of-Closest-Approach (PoCA) algorithm remains one of the most widely adopted methods due to its computational simplicity, robustness, and ability to rapidly localize scattering centers from reconstructed muon tracks \cite{10.1063/5.0273072,Schultz:2004kx,Tripathy:2018crw}. However, the resulting PoCA point clouds are typically affected by finite detector resolution, track-fitting uncertainties, and multiple-scattering fluctuations, producing diffuse object boundaries and a significant population of background points that complicate object reconstruction and material identification \cite{CAI2024169616,Jonkmans:2012sf}.

To address these limitations, recent developments in muon tomography have increasingly incorporated automated clustering and pattern-recognition techniques for object localization and noise suppression \cite{CAI2024169616,Jonkmans:2012sf,HOU20212348}. In realistic inspection scenarios, neither the number nor the positions of hidden objects are known \textit{a priori}. Density-based clustering algorithms provide an attractive solution because they can identify localized concentrations of high-scattering PoCA points while simultaneously rejecting sparse background events as noise. In this work, the Hierarchical Density-Based Spatial Clustering of Applications with Noise (HDBSCAN) algorithm \cite{campello2013density} is employed to automatically separate reconstructed PoCA clouds into individual object candidates. Unlike conventional clustering techniques, HDBSCAN does not require the number of clusters to be specified beforehand and can identify clusters with arbitrary geometries, making it particularly suitable for the reconstruction of multiple concealed objects within large interrogation volumes \cite{campello2013density,mcinnes2017hdbscan}.

A further challenge in material characterization arises from the broad momentum spectrum of atmospheric muons. Since the magnitude of multiple Coulomb scattering is inversely proportional to the muon momentum, low-momentum muons undergo substantially larger deflections than high-momentum muons traversing identical materials. Consequently, scattering-angle measurements alone provide limited discrimination between different materials and introduce significant event-by-event fluctuations. To overcome this limitation, the detector system incorporates a downstream magnetic spectrometer capable of estimating the momentum of individual muons from their magnetic deflection. The reconstructed momentum is subsequently combined with the measured scattering angle to define a momentum-corrected scattering density observable, enabling a more direct characterization of material properties while significantly improving the separation of low-$Z$ and high-$Z$ materials.

The momentum and path length corrected scattering density observable is given as
\begin{equation}
\rho_s =
\frac{(\theta p)^2}
{{L_{\mathrm{eff}}}},
\end{equation}

where $\theta$ denotes the reconstructed multiple-scattering angle, $p$ is the reconstructed muon momentum, and $L_{\mathrm{eff}}$ is the effective distance traveled by the muon inside the reconstructed object volume. Since the magnitude of multiple Coulomb scattering depends on both the material properties and the traversed thickness, the inclusion of $L_{\mathrm{eff}}$ provides a correction for differences in path length among individual muon trajectories. Consequently, $\rho_s$ becomes more sensitive to the intrinsic scattering characteristics of the material rather than to purely geometrical effects. This quantity is therefore adopted as the principal scattering-density observable used for material discrimination and object classification throughout this study.

In this study, a complete cosmic muon tomography system is designed and simulated using the GEANT4 framework \cite{AGOSTINELLI2003250}. The detector consists of six scintillator strip tracking stations arranged around the interrogation volume and a downstream four station magnetic spectrometer for momentum reconstruction. The tracking stations measure the incoming and outgoing muon trajectories, while the magnetic spectrometer provides event by event momentum estimates through magnetic deflection measurements. These reconstructed quantities are subsequently used to determine the scattering angle and the momentum-corrected scattering observables required for material characterization \cite{HIGHLAND1975497,Olive_2014}.

The reconstructed muon trajectories are employed in a PoCA reconstruction algorithm to estimate the most probable scattering locations within the interrogation volume. The resulting three-dimensional PoCA cloud is analyzed using the HDBSCAN algorithm, which identifies localized regions of enhanced scattering corresponding to hidden objects while suppressing background noise \cite{mcinnes2017hdbscan}. Geometrical and scattering related features are then extracted from each cluster to determine object locations and perform material discrimination.

The overall objective of this work is to develop an integrated muon tomography framework capable of simultaneous object localization, multi object detection, material identification. By combining momentum aware scattering measurements, PoCA reconstruction, density based clustering, and three dimensional image reconstruction within a unified analysis chain, the proposed system provides a complete end to end solution for cosmic muon tomography and material characterization.

\section{Muon Spectrum}
\label{sec:muonspectrum}

Cosmic ray muons are naturally occurring charged particles produced in the Earth's atmosphere through interactions of high energy primary cosmic rays, predominantly protons and heavier nuclei, with atmospheric nuclei. These interactions generate hadronic cascades, producing secondary pions ($\pi^{\pm}$) and kaons ($K^{\pm}$), which subsequently decay into muons. Owing to their relativistic velocities and mean lifetime of approximately $2.2~\mu$s, a significant fraction of these muons reach the Earth's surface before decaying~\cite{Gaisser:2016uoy,ParticleDataGroup:2024cfk}. Their high penetrating power makes them particularly useful for non-destructive imaging and muon tomography~\cite{schultz2007,MORRIS15102008}. The angular distribution of atmospheric muons is approximately described by

\begin{equation}
I(\theta)=I_0\cos^n\theta,
\end{equation}

where $I_0$ is the vertical muon intensity, $\theta$ is the zenith angle, and $n\approx 2$ for near-vertical directions~\cite{BAHMANABADI20191,ParticleDataGroup:2024cfk}. The integrated vertical muon flux at sea level is approximately $\Phi\approx 1~\mathrm{cm^{-2},min^{-1}}$, providing a continuous, naturally available source of penetrating particles for passive imaging applications without requiring artificial radiation sources.

\begin{figure}
    \centering
    \includegraphics[width=0.9\linewidth]{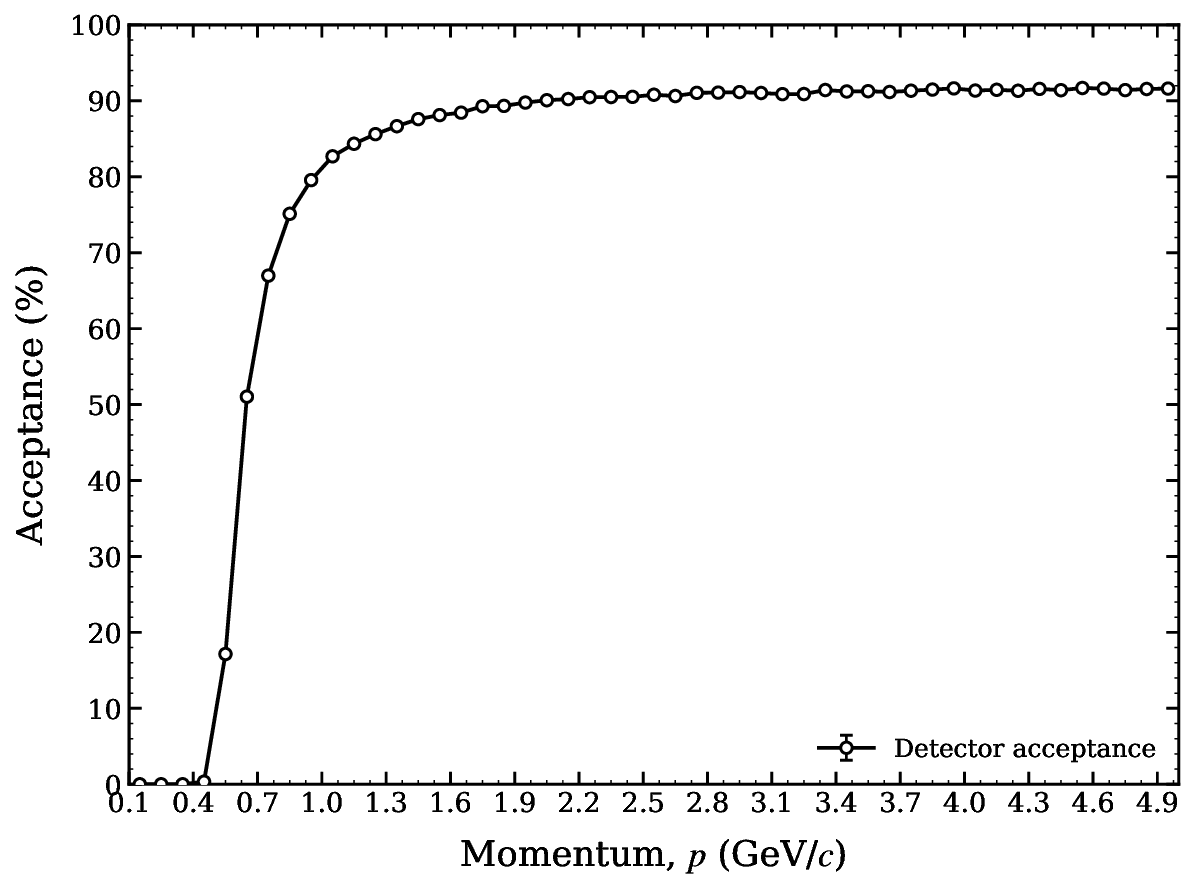}
    \caption{Muon momentum acceptance of the detector as a function of momentum for uniformly generated muons in the range of 0.1–5$~\mathrm{GeV}/c$.}
    \label{fig:Acceptance}
\end{figure}

To determine the muon momentum acceptance of the detector, an initial acceptance study was performed by uniformly generating muons over the momentum range of 0.1 – 5$~\mathrm{GeV}/c$ and propagating them through the detector geometry described in Section ~\ref{sec:detector}. Due to the presence of the magnetic field, low-momentum muons undergo significant bending and may fail to traverse the full detector geometry, resulting in reduced detector acceptance. Figure \ref{fig:Acceptance} shows the muon acceptance as a function of momentum over the investigated range. Based on this study, the lower momentum threshold for the generated muons was set to 0.5$~\mathrm{GeV}/c$, above which the muons have sufficient momentum to traverse the detector geometry. The upper momentum threshold was fixed at 10$~\mathrm{GeV}/c$, as at higher momenta the available statistics in this region are relatively limited. Thus, the final muon momentum range was chosen to be $0.5~\mathrm{GeV}/c \leq p \leq 10~\mathrm{GeV}/c$. This momentum interval contains the majority of atmospheric cosmic muons relevant for small-scale muon tomography while providing sufficient penetration through dense materials together with measurable multiple Coulomb scattering for material discrimination.

The generated muon momentum distribution is modelled using an approximate power law parameterization of the cosmic ray muon spectrum. This parameterization does not aim to reproduce the experimentally measured cosmic ray muon momentum distribution with high precision. Since the primary objective of this study is to evaluate the momentum measurement performance of the proposed detector, rather than to determine the absolute cosmic ray muon flux, an approximate momentum distribution is sufficient. The momentum dependence is accounted for through the momentum measurement performed by the detector. Therefore, this simplified power law distribution is adopted for generating muons in the simulation.

\begin{equation}
\label{equation_1}
\frac{d^2N}{dp\,d\Omega}
\propto
p^{-\gamma} f(\theta),
\end{equation}
where the angular dependence can be approximated as
\begin{equation}
f(\theta) \simeq \cos^2\theta.
\end{equation}

In this study, we adopt $\gamma=2.7$ ~\cite{https://doi.org/10.1029/92JA02672,HEBBEKER2002107} as an appropriate approximation for the momentum range considered. Here, $\theta$ denotes the zenith angle. The resulting power law distribution produces a substantially larger population of low momentum muons compared to high momentum muons.

\begin{figure}
\centering
\includegraphics[width=0.9\linewidth]{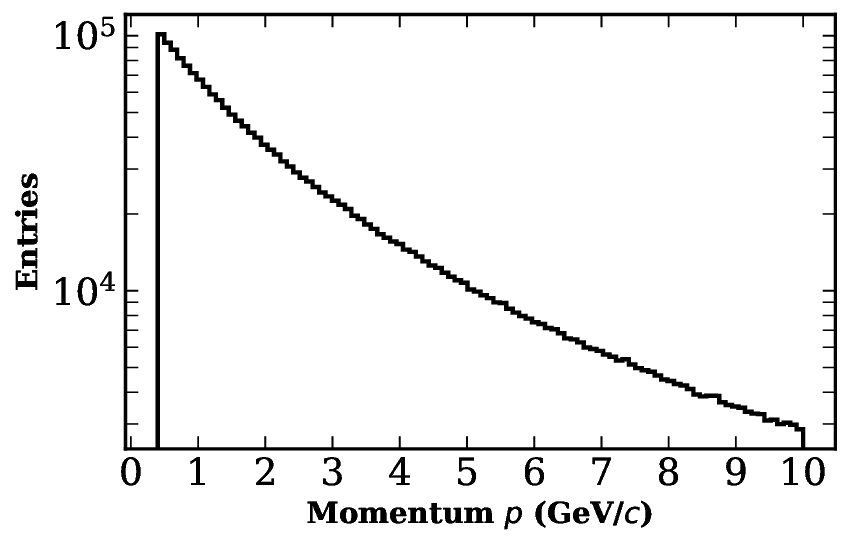}
\caption{Simulated cosmic-ray muon momentum spectrum used in this work. Muons were generated within the range of 0.5--10~GeV/$c$ following the measured atmospheric cosmic muon distribution.}
\label{fig:MuonSpectrum}
\end{figure}

The complete muon momentum spectrum used in the simulation is shown in Fig.~\ref{fig:MuonSpectrum}. The figure demonstrates the expected decrease in muon intensity with increasing momentum and illustrates the dominance of low momentum muons within the selected momentum range. The generated distribution serves as the primary input source for detector simulations, momentum reconstruction studies, and tomography analyses presented in this work.

\section{Detector Geometry Simulation}
\label{sec:detector}
\begin{figure}
\centering
\includegraphics[width=0.40\textwidth]{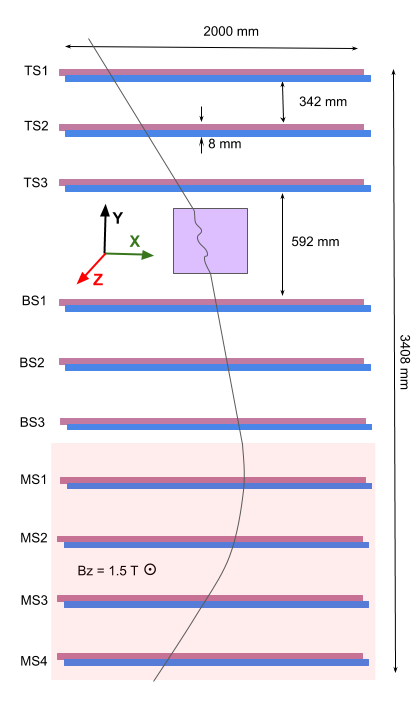}
\caption{
Schematic representation of the complete detector geometry showing the six tracking stations used for muon tomography, the interrogation volume, and the downstream magnetic spectrometer employed for momentum reconstruction.
}
\label{fig:detector_geometry}
\end{figure}

A dedicated cosmic ray muon tomography system was designed and implemented within the GEANT4 simulation framework~\cite{AGOSTINELLI2003250,ALLISON2016186}. The detector consists of two main subsystems: a six-station tracking system for muon tomography and a four-station magnetic spectrometer for muon momentum estimation. The six tracking stations are arranged around the interrogation volume, with three above and three below the target region. These stations are used to reconstruct the incoming and outgoing muon trajectories and to determine the scattering angle produced by interactions inside the inspected objects, following the standard multiple scattering muon tomography approach~\cite{borozdin2003,schultz2007,MORRIS15102008}. Below the tomography system, an additional four tracking stations are placed inside a uniform magnetic field region. These stations form a magnetic spectrometer that measures the curvature of the muon trajectory, allowing the momentum of each muon to be estimated on an event by event basis. The complete detector geometry is shown in Fig.~\ref{fig:detector_geometry}. The combined tracking and spectrometer configuration provides precise scattering-angle measurement, and momentum estimation, which are essential for accurate object localization, material identification, and three-dimensional image reconstruction.

\begin{figure}
    \centering
    \includegraphics[width=0.9\linewidth]{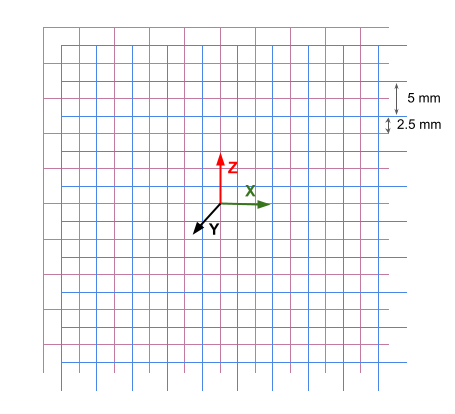}
    \caption{Transverse view of two detector layers with scintillator strips. The second layer (Blue) is shifted by $2.5$ mm with respect to the normal layer (Brown), implementing a staggered geometry that improves the effective hit position resolution.}
    \label{fig:layer_geometry}
\end{figure}

Each tracking plane consists of $400$ plastic scintillator strips with dimensions of $2000~\mathrm{mm} \times 5~\mathrm{mm} \times 2~\mathrm{mm}$, corresponding to the strip length, width, and thickness, respectively. The strips are arranged side-by-side to form an active detection area of approximately $2000~\mathrm{mm} \times 2000~\mathrm{mm}$. A single scintillator plane provides position information along one transverse coordinate. To obtain two-dimensional hit information, a second scintillator plane with identical dimensions is placed immediately below the first plane and rotated by $90^\circ$. In this configuration, the upper plane measures the $X$ coordinate while the lower plane measures the $Z$ coordinate of the traversing muon. The combination of these two orthogonal planes forms a detector layer capable of reconstructing the transverse position of the particle. The intrinsic strip pitch of a single plane is $5$ mm. To improve the position resolution without reducing the strip width, an additional orthogonal layer pair is installed directly below the first pair. This second pair is shifted by $2.5$ mm along both the $X$ and $Z$ directions, producing a staggered geometry. Figure~\ref{fig:layer_geometry} illustrate the staggered detector geometry used to achieve the improved spatial resolution. As a result, the effective detector pitch is reduced to $p_{\mathrm{eff}} = 2.5~\mathrm{mm}$. Assuming a uniform hit probability within the effective pitch, the corresponding spatial resolution is ~\cite{Chen_2023}

\begin{equation}
\sigma = \frac{p_{\mathrm{eff}}}{\sqrt{12}}
       \approx 0.72~\mathrm{mm}.
\end{equation}

\begin{figure}
    \centering
    \includegraphics[width=0.9\linewidth]{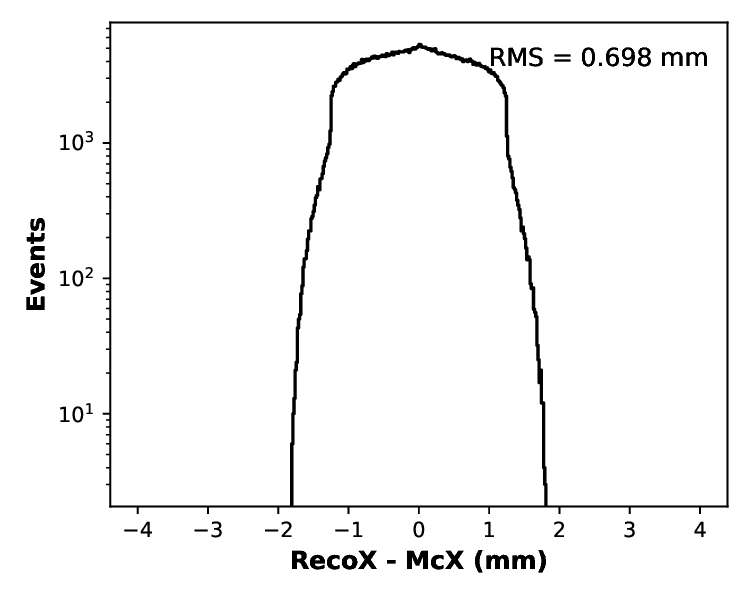}
    \includegraphics[width=0.9\linewidth]{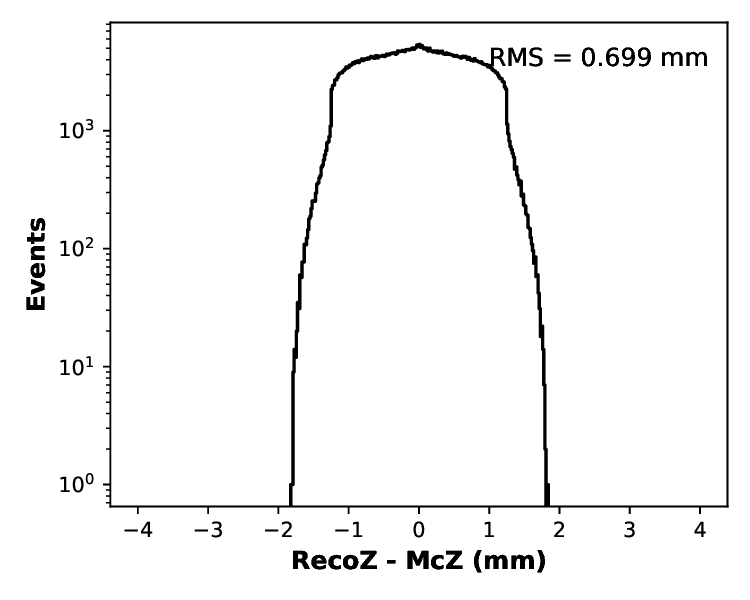}
    \caption{Residual distributions for detector layer along the $x$ direction (top panel) and $z$ direction (bottom panel).}
    \label{fig:resolution}
\end{figure}

The $2.5$ mm stagger between adjacent strip layers improves the detector position resolution. The detector resolution is estimated from the residual distribution, defined as the difference between the reconstructed hit position and the corresponding Monte Carlo truth position, $(x_{\mathrm{reco}} - x_{\mathrm{MC}})$. The width of the residual distribution is taken as the detector resolution. The residual distributions and the resulting resolutions for all detector layers are shown in Fig.~\ref{fig:resolution}.

One of the major limitations of conventional muon tomography systems is the absence of event by event momentum information. Since the multiple Coulomb scattering angle is inversely proportional to the particle momentum, variations in the cosmic muon momentum  can introduce significant uncertainties in material identification and density estimation. To overcome this limitation, a dedicated magnetic spectrometer was incorporated below the tomography detector to estimate the momentum of each individual muon before it is used in the tomography reconstruction. As shown in Fig. ~\ref{fig:detector_geometry}, the shaded region represents the magnetic field volume of the spectrometer, where a uniform magnetic field of $1.5$ T is applied to bend the muon trajectories for momentum measurement. As charged muons traverse this region, their trajectories are deflected by the Lorentz force, with the magnitude of the deflection depending on the particle momentum. In principle, the muon momentum can be estimated from the curvature of its trajectory in the magnetic field. However, instead of relying on an analytical curvature-based reconstruction, this work employs a data driven approach based on machine learning.

For the detector simulation, the muon trajectory is reconstructed on an event by event basis using the hit positions recorded by the four spectrometer tracking stations. A track fit is then performed to extract the relevant trajectory parameters. Rather than employing a dedicated global track reconstruction framework, the fitted track parameters and magnetic deflection observables are used directly as input features for the machine learning model. The fitted track parameters and magnetic deflection observables are then used as input features to a machine learning regression model trained to predict the muon momentum. This approach provides a robust momentum estimation over the full cosmic muon momentum range considered in this study while naturally incorporating detector resolution effects, multiple scattering, and measurement uncertainties.

\section{Hit Reconstruction}
\begin{figure}
    \centering
    \includegraphics[width=0.9\linewidth]{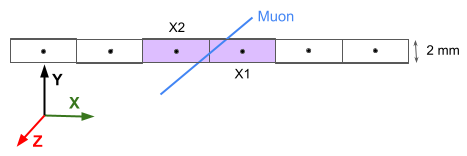}
    \caption{Schematic representation of a muon crossing two adjacent scintillator strips.}
    \label{fig:centroid}
\end{figure}

The detector simulation produces strip level hit information for each traversing cosmic ray muon. Since a single muon can produce signals in multiple neighboring scintillator strips, a dedicated reconstruction pipeline was developed to convert the raw detector response into precise three dimensional space points suitable for track reconstruction, momentum estimation, and muon tomography. As illustrated in Fig.~\ref{fig:centroid}, when a muon traverses the boundary between two adjacent strips, both strips may register hits. In such cases, the reconstructed hit position is assigned to the geometric midpoint of the two strip centers, providing a single representative space point for the detector layer. This simple geometric averaging improves the effective spatial resolution compared to assigning the hit to the center of a single strip. The centroid hit position is given as:

\begin{equation}
x_{\mathrm{centroid}}
=
\frac{1}{n}\sum_{i=1}^{n} x_i,
\qquad
z_{\mathrm{centroid}}
=
\frac{1}{n}\sum_{i=1}^{n} z_i.
\end{equation}
where $x_i$ and $z_i$ denotes the center position of strip $i$. The same procedure is applied independently to all detector planes. 

 After centroid reconstruction, the information from the two planes is combined to determine the full three dimensional hit position. The reconstructed $x$ coordinate is obtained from the first strip plane, the reconstructed $z$ coordinate is obtained from the second strip plane, and the $y$ coordinate is assigned according to the known geometric position of the detector layer. This procedure transforms two independent one dimensional measurements into a complete spatial measurement point. Each detector station consists of two staggered detector layers. The hit positions reconstructed in the two layers are combined by taking their average, resulting in a single station coordinate with improved spatial resolution. The improvement in resolution achieved by the staggered geometry is shown in Fig.~\ref{fig:resolution}. The resulting station coordinate provides the best estimate of the muon position at that measurement plane. The reconstructed station coordinates constitute the fundamental input for the remainder of the analysis.

The reconstructed detector hit positions are processed using a least squares straight line fitting procedure to determine the muon trajectory. For each event, the three tracking stations located above the interrogation volume are used to reconstruct the incoming track, while the three stations below the interrogation volume are used to reconstruct the outgoing track. The fitting procedure determines the best straight line passing through the measured hit positions in both the $x$-$y$ and $z$-$y$ projections. From the fitted slopes, a normalized three dimensional direction vector is obtained for each track.

\begin{figure}
    \centering
    \includegraphics[width=0.9\linewidth]{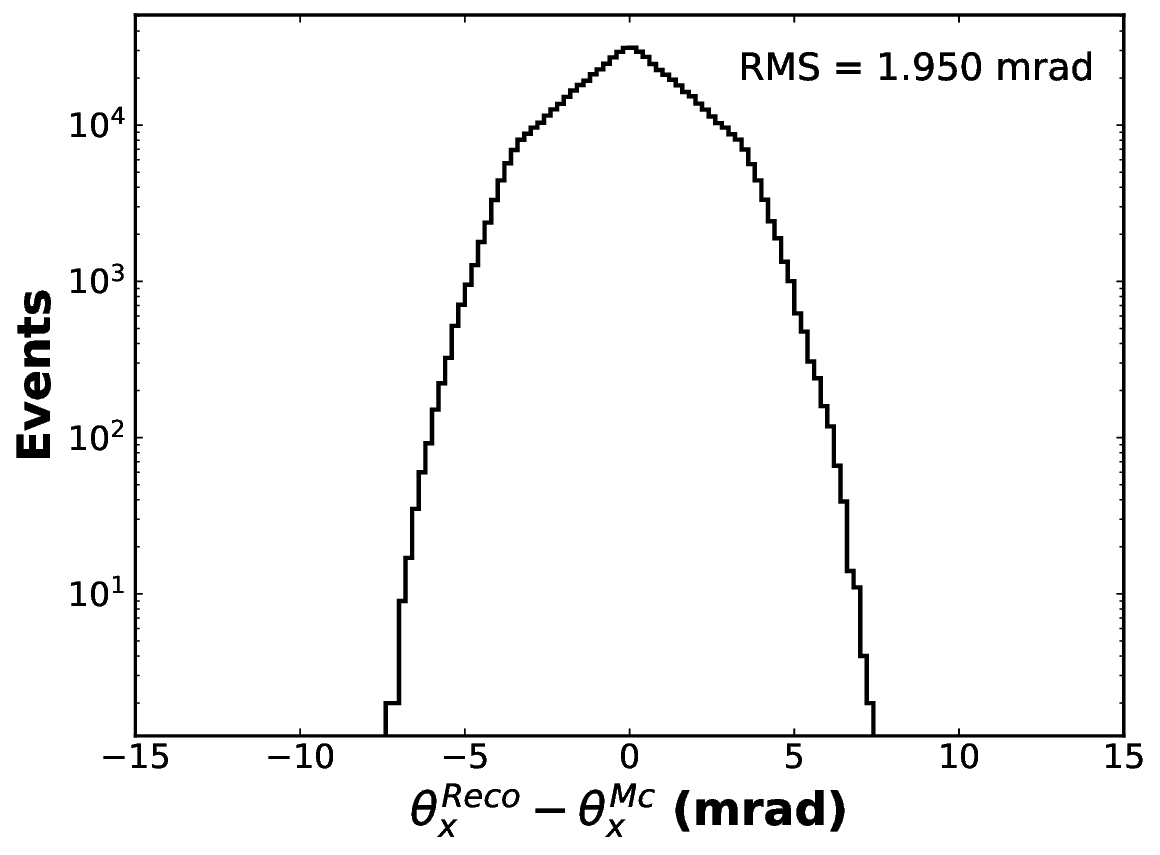}
    \includegraphics[width=0.9\linewidth]{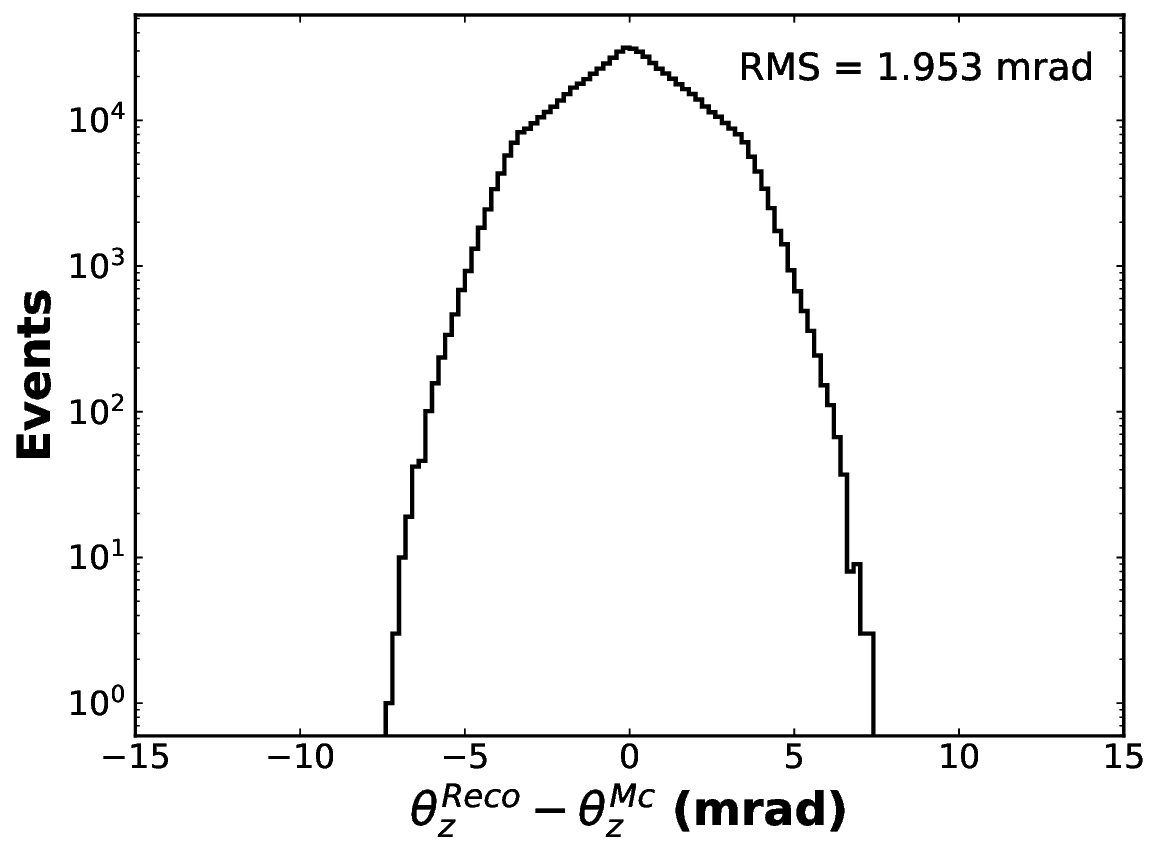}
    \caption{Detector angular resolution distribution for $x$-projection (top panel) and $z$-projection (bottom panel).}
    \label{fig:angular_resolution_xz}
\end{figure}

The incoming and outgoing track directions are then compared to calculate the muon scattering angle produced inside the interrogation volume. The scattering is evaluated separately in the $x$ and $z$ projections and subsequently combined to obtain the total scattering angle. The performance of the track reconstruction algorithm is evaluated using Monte Carlo truth information. The reconstructed scattering angles are compared with the corresponding true scattering angles, and the differences are used to determine the detector angular resolution. Resolution distributions are produced separately for the $x$-projection and the $z$-projection. The corresponding angular resolution distributions are shown in Fig.~\ref{fig:angular_resolution_xz}.

In addition to the tomography tracking stations, four detector stations located inside the magnetic field region are used for momentum estimation. The incoming muon direction is first determined using the last tomography tracking station (BS3), which serves as the reference direction before the magnetic field. Subsequently, five angular observables, denoted as $\alpha_1$, $\alpha_2$, $\alpha_3$, $\alpha_4$, and $\alpha_5$, are calculated from the directions measured at BS3 and the four magnetic spectrometer stations (MS1--MS4). Here, $\alpha_1$ corresponds to the angle between the incoming track at BS3 and the direction toward MS1, while $\alpha_2$, $\alpha_3$, $\alpha_4$, and $\alpha_5$ are defined similarly using the directions toward MS2, MS3, MS4, and the reconstructed downstream trajectory, respectively. Each angular observable is computed as the angle of the corresponding track segment with respect to the horizontal axis in the bending plane. Together, the set of angular variables $(\alpha_1,\alpha_2,\alpha_3,\alpha_4,\alpha_5)$ characterizes the magnetic deflection of the muon and forms the primary input to the momentum reconstruction algorithm.

\section{ML Approch For Muon Momentum Prediction}

Accurate estimation of the incoming muon momentum is essential for enhancing the material discrimination capability of a muon tomography system. The multiple Coulomb scattering angle experienced by a muon traversing matter is strongly dependent on its momentum, with lower momentum muons exhibiting larger angular deviations. Since cosmic muons arrive with a broad momentum spectrum and no direct momentum measurement is available in conventional tomography systems, a dedicated momentum reconstruction framework has been developed using additional tracking stations placed below the detector volume.

In this approach, the trajectory bending induced by the magnetic field is quantified through a set of deflection angles extracted from reconstructed muon tracks. These angular observables are subsequently used as input features to a machine learning regression model designed to estimate the momentum of individual muons. A dedicated Monte Carlo simulation dataset was generated using the detector geometry described in the previous section. Cosmic muons were produced according to the momentum spectrum defined in Section~\ref{sec:muonspectrum}, covering the range from $0.5$~GeV/$c$ to $10$~GeV/$c$. For each simulated event, the true momentum components $(p_x, p_y, p_z)$ were recorded, along with the corresponding reconstructed bending angles measured in the magnetic spectrometer region.

The resulting dataset consisted of reconstructed angular observables together with the corresponding true momentum values obtained from the Monte Carlo simulation. Five bending angles, denoted as $\alpha_1,\alpha_2,\alpha_3,\alpha_4,\alpha_5$ were extracted from the last $5$ layers of the detector. These angles describe the evolution of the muon trajectory through the magnetic spectrometer. To enhance the sensitivity to momentum-dependent curvature, additional derived features were constructed from the angular measurements. To characterize the change in the muon trajectory within the magnetic spectrometer, the angular differences were computed by taking the first angular measurement, $\alpha_1$, as the reference. The relative angular variables are defined as

\begin{align}
\Delta\alpha_{21} &= \alpha_2-\alpha_1,\\
\Delta\alpha_{31} &= \alpha_3-\alpha_1,\\
\Delta\alpha_{41} &= \alpha_4-\alpha_1,\\
\Delta\alpha_{51} &= \alpha_5-\alpha_1.
\end{align}

These variables provide information about cumulative trajectory bending and therefore improve the model's ability to predict the total momentum of the muon.

Muon momentum prediction was formulated as a supervised regression problem. A Random Forest Regressor was selected due to its robustness against non linear feature correlations and its ability to model complex relationships without requiring extensive parameter tuning. The model consists of an ensemble of decision trees whose predictions are averaged to obtain the final momentum estimate. For the training of the regression model, a sample of $1.2\times10^{6}$ simulated muons was used, covering the momentum range from $0.5$ to $10~\mathrm{GeV}/c$. This large training sample was used to provide sufficient statistics across the considered momentum range and to ensure that the model could learn the relationship between the detector observables and the true muon momentum. 

\begin{figure}
    \centering
    \includegraphics[width=0.9\linewidth]{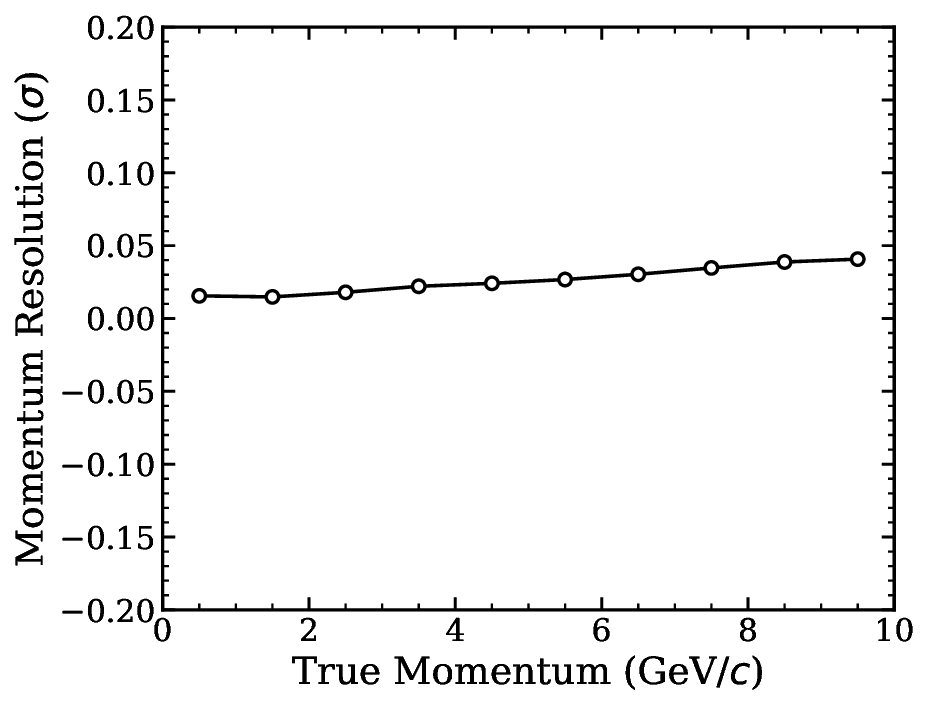}
    \caption{Relative momentum resolution as a function of the true muon momentum.}
    \label{fig:Momentum_resolution1}
\end{figure}

The Random Forest model consisting of 200 decision trees, with a maximum tree depth of 20. The minimum number of samples required to split an internal node and form a leaf node is set to 5 and 2, respectively. The performance of the trained model is evaluated in terms of the relative momentum resolution as a function of the true muon momentum, as shown in Fig.~\ref{fig:Momentum_resolution1}. The momentum resolution remains below approximately 5\% over the full momentum range considered. The training and testing performance metrics, summarized in Table~\ref{tab:rf_performance}, show close agreement, indicating good model generalization. This demonstrates that the proposed spectrometer geometry, combined with the Random Forest regression model, provides accurate and stable momentum reconstruction for cosmic muons, making it suitable for scattering based material identification in the tomography system.

\begin{table}[htbp]
\centering
\caption{Performance of the Random Forest regression model evaluated on the training and testing datasets.}
\label{tab:rf_performance}
\begin{tabular}{lcc}
\hline
\textbf{Performance Metric} & \textbf{Training} & \textbf{Testing} \\
\hline
MAE (GeV)          & 0.052   & 0.058 \\
RMSE (GeV)         & 0.105   & 0.109 \\
$R^2$ Score        & 0.998  & 0.997 \\
Relative Bias (\%) & 0.0042  & 0.060 \\
\hline
\end{tabular}
\end{table}

\begin{figure}
    \centering
    \includegraphics[width=0.9\linewidth]{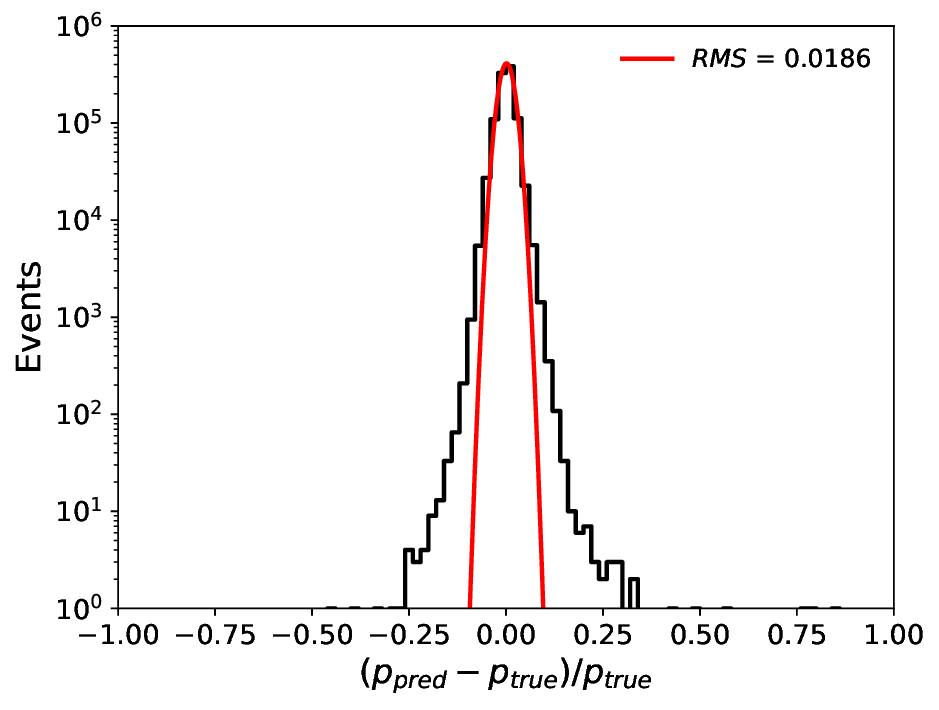}
    \caption{Relative momentum resolution distribution along with the corresponding Gaussian fit.}
    \label{fig:Momentum_resolution}
\end{figure}

Figure~\ref{fig:Momentum_resolution} shows the distribution of the relative momentum resolution for a representative momentum sample together with the corresponding Gaussian fit. The fitted width of the distribution is used to determine the momentum resolution, while the mean indicates the reconstruction bias. The narrow Gaussian distribution centered close to zero demonstrates that the proposed momentum reconstruction method achieves both high precision and negligible systematic bias.

The trained model was subsequently applied to all reconstructed tomography events to estimate the momentum of individual cosmic muons. These momentum estimates were then incorporated into the scattering density calculation used for material identification and three dimensional image reconstruction.

\section{PoCA Based Material Identification}

The final objective of this work is the material identification, and three-dimensional reconstruction of the objects using cosmic muons. Following hit reconstruction, track fitting, scattering-angle estimation, and machine learning based momentum prediction, a complete muon tomography framework is implemented to image the contents of the interrogation volume.

The reconstructed incoming and outgoing muon trajectories are used to estimate the most probable scattering location within the target volume using the Point of Closest Approach (PoCA) algorithm. The incoming track is represented by a point $\mathbf{P}_{1}$ and a direction vector $\mathbf{V}_{1}$ obtained from the upper tracking stations, while the outgoing track is described by a point $\mathbf{P}_{2}$ and a direction vector $\mathbf{V}_{2}$ reconstructed from the lower tracking stations.

The coordinates $(x_{\mathrm{PoCA}},\, y_{\mathrm{PoCA}},\, z_{\mathrm{PoCA}})$ define the reconstructed interaction point of the muon within the tomography volume. For each reconstructed PoCA point, the measured scattering angle and the predicted muon momentum are combined to calculate a scattering density parameter. Since the magnitude of multiple Coulomb scattering depends on both the material properties and the muon momentum, the inclusion of momentum information significantly improves the discrimination between low, medium, and high $Z$ materials. The reconstructed PoCA points are subsequently accumulated in three dimensional voxels to generate a volumetric image of the interrogation region, enabling the localization and identification of objects.

\begin{figure}
\centering
\includegraphics[width=0.9\linewidth]{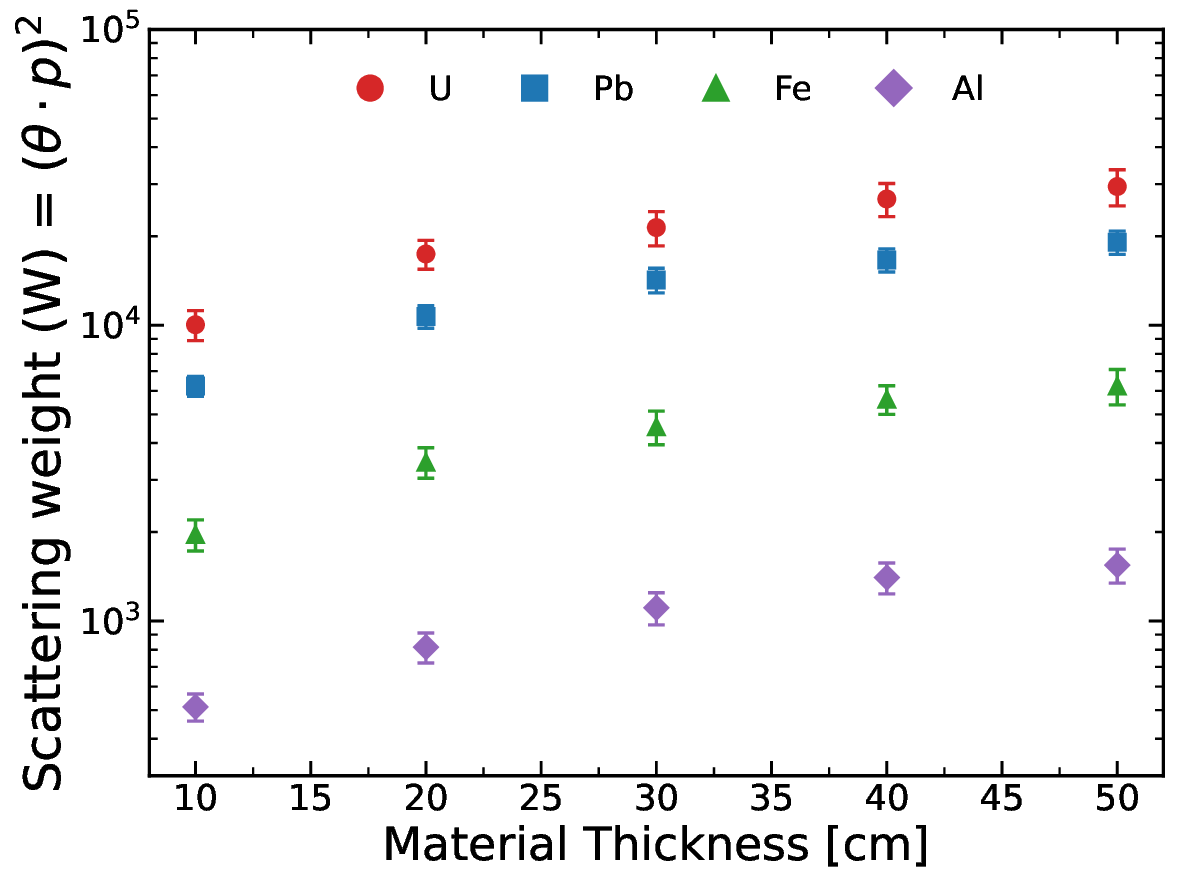}
\caption{Variation of the scattering weight, $W=(\theta p)^2$, as a function of material thickness for Al, Fe, Pb, and U.}
\label{fig:weight_thickness}
\end{figure}

The variation of the scattering weight,
\begin{equation}
W = (\theta p)^2,
\end{equation}
for different material thicknesses is shown in Fig.~\ref{fig:weight_thickness}. Although the momentum correction significantly reduces the dependence of the scattering observable on the incident muon momentum, the scattering weight still exhibits a strong dependence on the material thickness. As the thickness increases, the accumulated multiple Coulomb scattering also increases, resulting in larger values of $W$. Consequently, different materials with different thicknesses can have similar scattering weights, making it difficult to identify the material using only $W$. The amount of multiple Coulomb scattering experienced by a muon depends not only on its momentum but also on the distance traveled within the material. While low momentum muons undergo larger angular deflections than high momentum muons, a longer path length through the material also increases the total scattering. Therefore, the scattering weight alone cannot uniquely characterize the material. To minimize the dependence on both the muon momentum and the traversed path length, a momentum and path length normalized scattering density parameter is defined as

\begin{equation}
\label{eq:scattering_density}
\rho_s=\frac{(\theta p)^2}{L_{\mathrm{eff}}},
\end{equation}

where $\theta$ is the reconstructed total scattering angle, $p$ is the machine-learning-predicted muon momentum, and $L_{\mathrm{eff}}$ is the effective path length of the reconstructed muon trajectory inside the object. This normalization suppresses the thickness dependence of the scattering observable, allowing $\rho_s$ to be determined primarily by the intrinsic scattering properties of the material. As a result, the scattering density parameter provides improved separation between materials with different atomic numbers and densities, making it more suitable for material identification in cosmic muon tomography.

\begin{figure}
    \centering
    \includegraphics[width=0.9\linewidth]{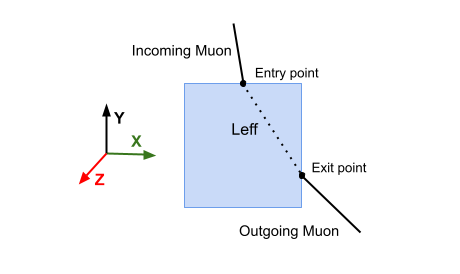}
    \caption{Schematic representation of the effective path length, $L_{\mathrm{eff}}$, used to describe the distance travelled by a particle through a material.}
    \label{fig:Leff_dig}
\end{figure}

The effective path length, $L_{\mathrm{eff}}$, is calculated using the object geometry obtained from the clustering stage. The object center and dimensions are estimated from the HDBSCAN clustering of the reconstructed PoCA points, which provides the reconstructed object position and size. A ray box intersection algorithm is employed to determine the entry and exit points of the muon track within the object. The effective path length, $L_{\mathrm{eff}}$, is then calculated as the Euclidean distance between these two intersection points, as illustrated in Fig.~\ref{fig:Leff_dig}.

\begin{figure}
    \centering
    \includegraphics[width=0.9\linewidth]{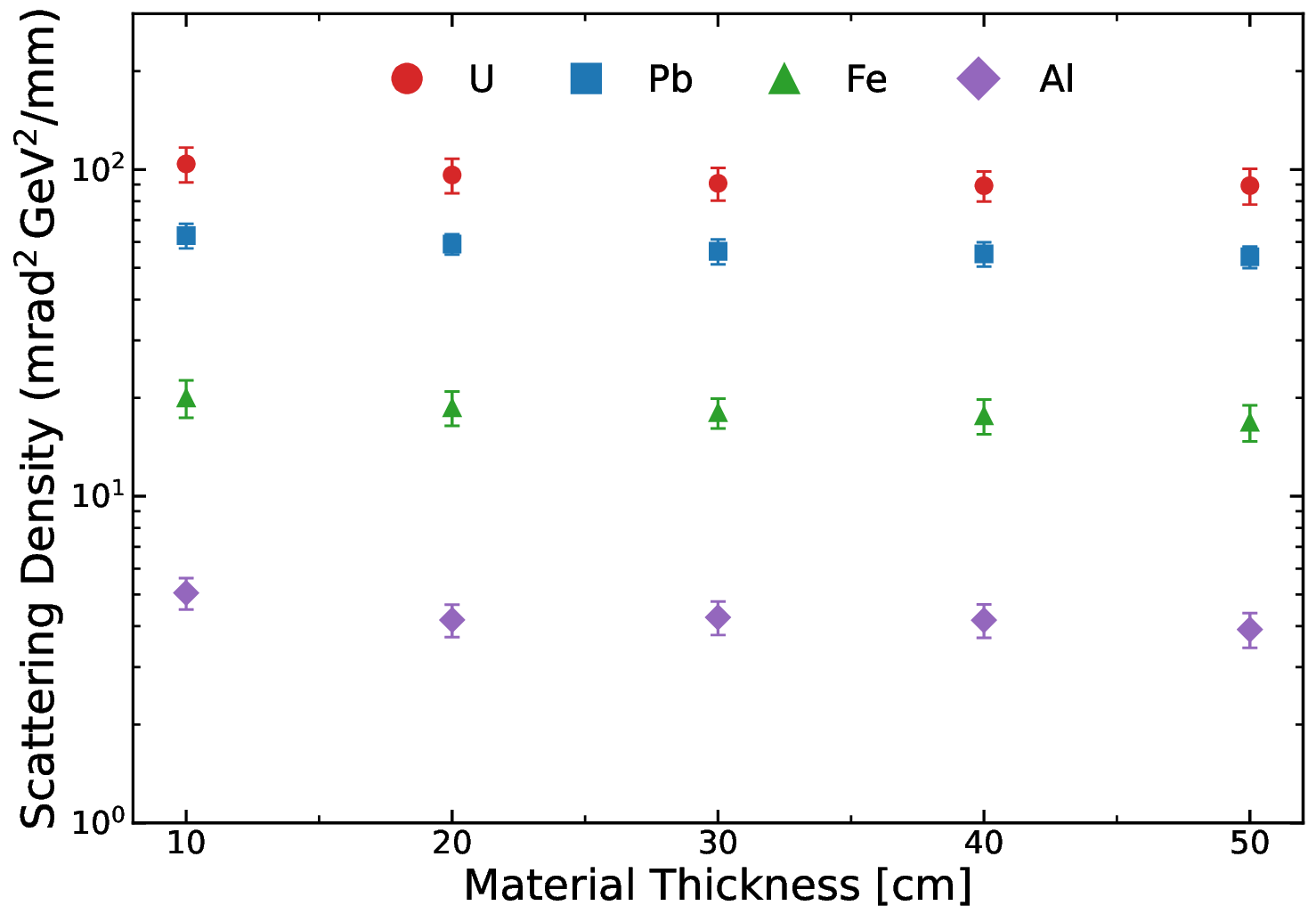}
    \caption{Reconstructed scattering density parameter as a function of material thickness for four different materials.}
    \label{fig:ScatteringDensityVsThickness}
\end{figure}

The parameter ($\rho_s$) is proportional to the scattering power of the traversed medium and provides a more robust observable for material characterization. By accounting for both momentum variations and differences in material thickness, $\rho_s$ enhances the separation between materials of different atomic composition and density. For the study presented in Fig.~\ref{fig:ScatteringDensityVsThickness}, the effective path length, $L_{\mathrm{eff}}$, is calculated using the true object geometry obtained from the simulation. This approach isolates the intrinsic behavior of the proposed scattering density parameter by eliminating uncertainties associated with object reconstruction, thereby allowing the dependence of $\rho_s$ on material type and thickness to be evaluated independently. In the complete reconstruction chain, however, $L_{\mathrm{eff}}$ is computed using the reconstructed object center and dimensions obtained from the HDBSCAN clustering of the reconstructed PoCA points.

Figure~\ref{fig:ScatteringDensityVsThickness} shows the variation of the reconstructed scattering density parameter as a function of material thickness for Al, Fe, Pb, and U. As expected, the scattering density increases with the atomic number of the material, reflecting the stronger multiple Coulomb scattering experienced by muons traversing high-$Z$ media. For each material, the reconstructed scattering density remains nearly constant over the investigated thickness range, indicating that the path length normalization (Eq.~\ref{eq:scattering_density}) effectively suppresses the thickness dependence of the scattering observable. A slight increase in the reconstructed scattering density is observed for thinner targets. This behavior arises because the effective path length, $L_{\mathrm{eff}}$, varies linearly with thickness, whereas the $(\theta * p)^2$ does not scale linearly over the entire thickness range as can seen in the figure ~\ref{fig:weight_thickness}. Consequently, the normalization by $L_{\mathrm{eff}}$ produces slightly larger values of the scattering density parameter for thinner objects. Despite this small trend, the separation between Al, Fe, Pb, and U remains clear, demonstrating that the scattering density estimator is primarily sensitive to the intrinsic material properties rather than the object thickness. In particular, high-$Z$ materials such as Pb and U exhibit consistently larger scattering densities than low-$Z$ materials, providing a robust basis for material identification in muon tomography applications.

\begin{figure}
    \centering
    \includegraphics[width=0.9\linewidth]{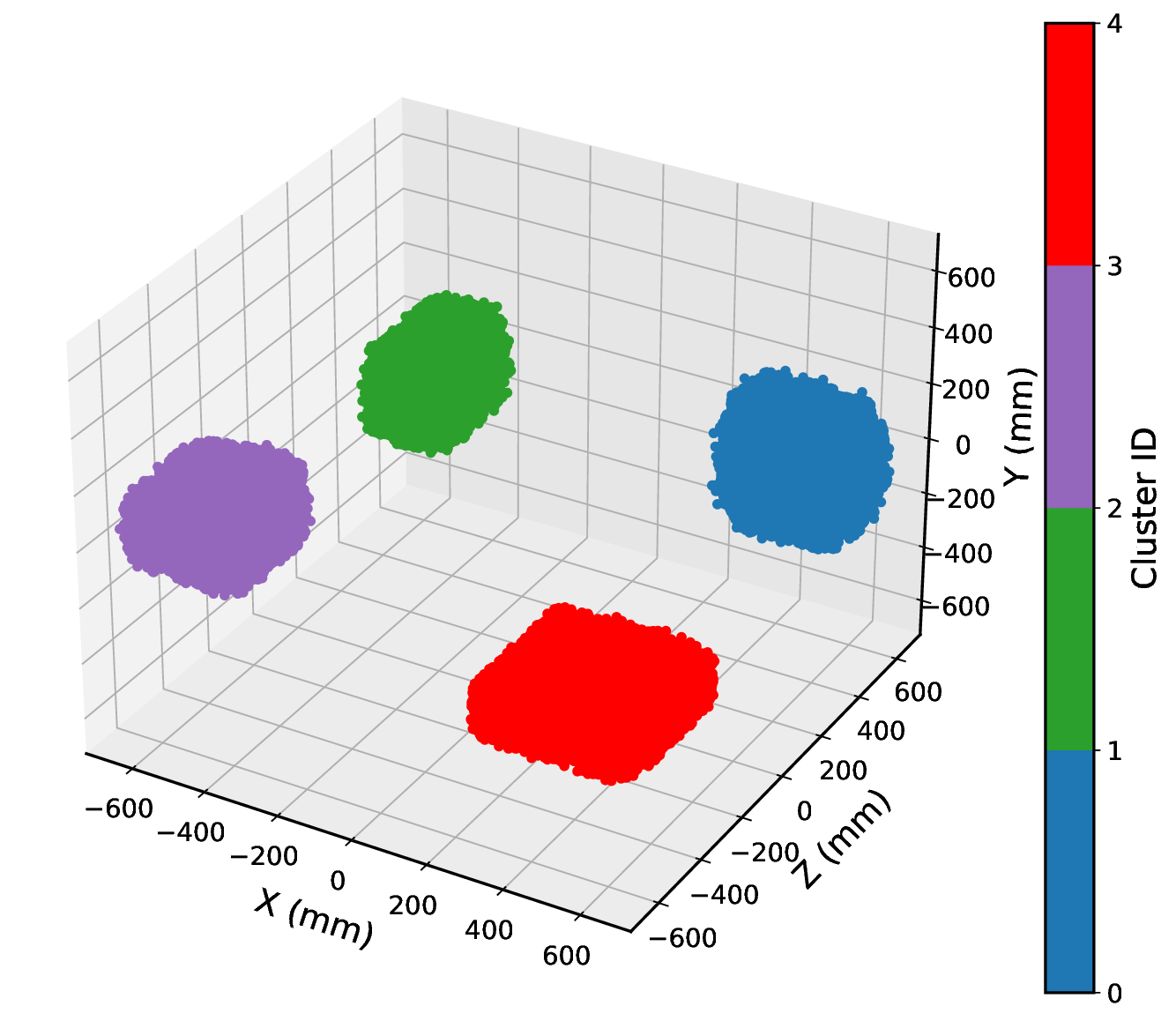}
    \caption{Three dimensional PoCA reconstruction}
    \label{fig:3D_clusters}
\end{figure}

After validating the scattering density formulation using materials of known thickness, the next challenge is its application to realistic scenarios where neither the material composition nor its dimensions are known \textit{a priori}. The primary objective of a muon tomography system is therefore not only to identify the material composition but also to localize the hidden object and reconstruct its three-dimensional geometry. To investigate this capability, multiple objects composed of different materials and having different dimensions were simultaneously placed inside the interrogation volume. A sample of $1.0\times10^{6}$ simulated cosmic muons was used for the material identification and object reconstruction studies. The large number of muons provides sufficient statistics for the reconstruction of the PoCA distribution and for reliable estimation of the scattering density. The resulting PoCA reconstruction is shown in Figure~\ref{fig:3D_clusters}.

For each muon event, a momentum corrected scattering weight is first calculated using the reconstructed scattering angle $\theta$ and the predicted momentum $p$. The collection of PoCA points forms a three dimensional point cloud that contains information about the location of scattering centers inside the volume. Regions containing dense materials generate localized concentrations of high scattering PoCA points, while the surrounding background produces sparse and diffuse distributions.

The first step in the reconstruction procedure is therefore to determine how many distinct objects are present inside the interrogation volume. For this purpose, the Hierarchical Density Based Spatial Clustering of Applications with Noise (HDBSCAN) algorithm is applied to the reconstructed PoCA point cloud. The HDBSCAN algorithm is particularly well suited for muon tomography because it identifies clusters based on local point density, can reconstruct clusters of arbitrary shape, and automatically rejects isolated background points as noise. Unlike conventional clustering techniques, it does not require the number of clusters to be specified beforehand, making it highly suitable for the analysis of unknown target configurations.


\begin{table*}[t]
\centering
\caption{Comparison between the true and reconstructed object centers and widths obtained from the HDBSCAN clustering analysis.}
\label{tab:reco_comparison}
\resizebox{2.1\columnwidth}{!}{
\begin{tabular}{lccccccc}
\hline\hline
Cluster &
\begin{tabular}[c]{@{}c@{}}
True Center \\
$(x,y,z)$
[cm]
\end{tabular} &
\begin{tabular}[c]{@{}c@{}}
Reco Center \\
$(x,y,z)$ [cm]
\end{tabular} &
\begin{tabular}[c]{@{}c@{}}
Position Error \\
$(\Delta x,\Delta y,\Delta z)$ [cm]
\end{tabular} &
\begin{tabular}[c]{@{}c@{}}
True Size \\
$(x,y,z)$[cm]
\end{tabular} &
\begin{tabular}[c]{@{}c@{}}
Reco Size \\
$(x,y,z)$ [cm]
\end{tabular} &
\begin{tabular}[c]{@{}c@{}}
Size Error \\
$(\Delta x,\Delta y,\Delta z)$ [cm]
\end{tabular} &
\begin{tabular}[c]{@{}c@{}}
Density \\
$Mean\pm \sigma$
\end{tabular} \\
\hline

0 &
$(50.0,0.0,50.0)$ &
$(49.9, -1.3, 49.9)$ &
$(0.1,1.3,0.1)$ &
$(30.0,50.0,20.0)$ &
$(31.9, 49.4, 23.6)$ &
$(1.9,0.6,3.6)$ & $3.47\pm 0.4$ \\

1 &
$(-50.0,-5.0,50.0)$ &
$(-49.8,-5.8,50.0)$ &
$(0.2,0.8,0.0)$ &
$(20.0,40.0,30.0)$ &
$(23.2,41.1,31.8)$ &
$(3.2,1.1,1.8)$ & $14.7 \pm 1.9$ \\

2 &
$(-50.0,10.0,-50.0)$ &
$(-49.5,7.4,-49.9)$ &
$(0.5,2.6,0.1)$ &
$(30.0,30.0,30.0)$ &
$(30.4,30.2,30.6)$ &
$(0.4,0.2,0.6)$ & $55.6\pm 6.9$ \\

3 &
$(50.0,-15.0,-50.0)$ &
$(50.0,-15.8,-50.1)$ &
$(0.0,0.8,0.1)$ &
$(40.0,20.0,40.0)$ &
$(39.7,19.5,39.6)$ &
$(0.3,0.5,0.4)$ & $106.8 \pm 13.3$ \\

\hline\hline
\end{tabular}
}
\end{table*}


The clustering process separates the PoCA distribution into distinct groups, each corresponding to an individual physical object. Once the clusters have been identified, the geometric properties of each reconstructed object are extracted. The centroid of every cluster is calculated to determine the object location, while the spatial extent of the cluster along the three coordinate axes is used to estimate its dimensions. The true and reconstructed properties of all the cluster are given in the Table ~\ref{tab:reco_comparison}. These reconstructed dimensions provide an estimate of the effective material thickness encountered by the muons.

Using the reconstructed geometry, the effective path length $L_{\mathrm{eff}}$ is determined for each muon track for each cluster. This reconstructed path length is then incorporated into the scattering density estimator in Eq.~\ref{eq:scattering_density} which removes the dominant dependence on material thickness and allows the scattering response to be compared between objects of different sizes. 

\begin{figure}
    \centering
    \includegraphics[width=0.9\linewidth]{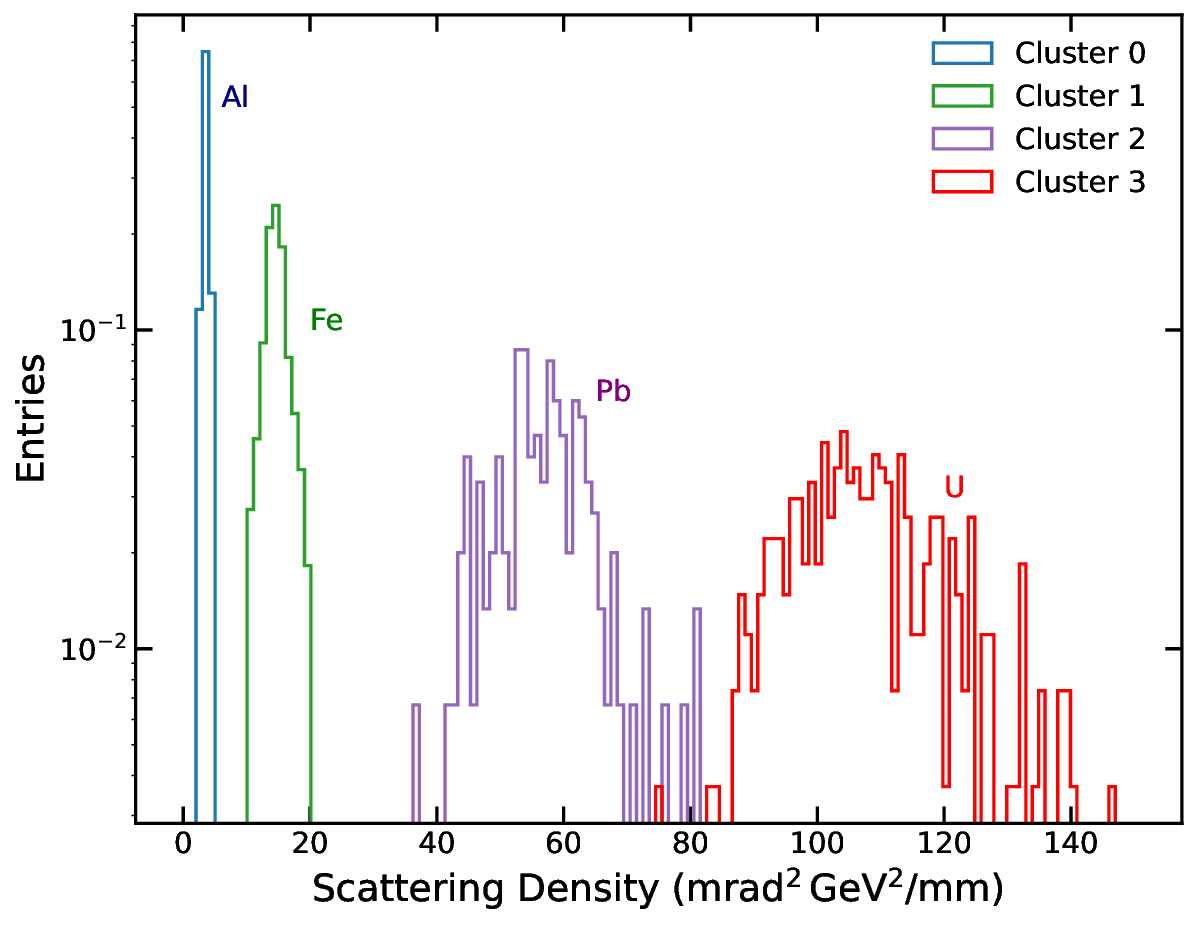}
    \caption{Distributions of the reconstructed scattering density for the different materials studied.}
    \label{fig:MaterialHisto}
\end{figure}

Because high-$Z$ materials produce larger multiple scattering angles than low-$Z$ materials, they generate systematically larger values of $\rho_s$. Consequently, clusters exhibiting larger mean scattering densities correspond to materials with higher atomic numbers. The reconstructed scattering density distributions for the identified clusters are shown in Figure~\ref{fig:MaterialHisto}. Despite differences in object shape and dimensions, a clear separation between the scattering density distributions of different materials is observed. The extracted scattering features therefore provide a robust basis for material discrimination. By combining object localization, geometric reconstruction, effective path length estimation, and scattering density analysis, the proposed framework enables the simultaneous detection, reconstruction, and identification of multiple hidden materials within the interrogation volume.\\
 
\section{Summary and Conclusions}

In this work, a complete cosmic muon tomography framework has been developed for the detection, localization, material identification, and three dimensional reconstruction of objects. The study combines detector hit reconstruction, scattering-angle determination, momentum prediction, PoCA reconstruction, and density based clustering into a unified imaging pipeline. The reconstruction procedure begins with the determination of incoming and outgoing muon trajectories using the tracking detector system. The Point of Closest Approach (PoCA) algorithm is then employed to estimate the most probable scattering location for each muon event. To remove the intrinsic momentum dependence of multiple Coulomb scattering, a momentum corrected scattering density parameter is introduced. Furthermore, the effective path length of each muon inside the reconstructed object is incorporated, allowing the scattering density to be expressed as a material dependent quantity that is largely independent of object thickness.

To evaluate the performance of the proposed methodology, multiple materials with different atomic numbers and dimensions were simultaneously placed inside the interrogation volume. The reconstructed PoCA distributions exhibited localized high density regions corresponding to the hidden objects. The HDBSCAN clustering algorithm was subsequently applied to automatically identify these regions without requiring prior knowledge of the number of objects. HDBSCAN successfully separated the reconstructed PoCA cloud into individual clusters while rejecting sparse background points as noise. Following cluster identification, the geometrical properties of each object were extracted. The cluster centers were reconstructed with sub centimeter accuracy in the transverse directions and within a few centimeters along the vertical direction. The reconstructed dimensions showed good agreement with the true object sizes, particularly for objects with moderate and large thicknesses. The comparison presented in Table~\ref{tab:reco_comparison} demonstrates that the localization errors remain below a few centimeters for all reconstructed objects, while the reconstructed widths closely reproduce the true dimensions.

The scattering density analysis further enabled material discrimination. The reconstructed density values increased systematically with atomic number, allowing clear separation between low-$Z$ and high-$Z$ materials. The measured mean scattering densities were found to follow the expected ordering,
$\mathrm{Al} < \mathrm{Fe} < \mathrm{Pb} < \mathrm{U} $ demonstrating the capability of the proposed approach to identify material composition in addition to object geometry. The normalized scattering density distributions remained well separated even for objects of different dimensions, confirming the effectiveness of the path length correction procedure. Overall, the results demonstrate that the developed muon tomography framework can simultaneously reconstruct the position, size, and material properties of multiple objects within a interrogation volume. The combination of machine learning based momentum estimation, PoCA reconstruction, effective path length correction, and HDBSCAN clustering provides a robust methodology for passive imaging applications.

Future developments will focus on improving the reconstruction of thin objects, where PoCA smearing and multiple scattering fluctuations can lead to overestimation of object dimensions. Advanced reconstruction techniques based on voxelized likelihood methods, weighted density estimation, and iterative scattering reconstruction are expected to further improve spatial resolution and material classification performance. These developments will enhance the applicability of cosmic muon tomography to nuclear safeguards, cargo inspection, waste characterization, and non-destructive imaging of dense structures.

 \section{Acknowledgement}
Bharat Kumar Sirasva acknowledges the financial support provided by the University Grants Commission (UGC) through the research fellowship and the IISER Mohali for providing access to its high performance computing (HPC) facilities used in this work. Rohit Gupta acknowledges the One Nation One Subscription (ONOS) initiative of the Government of India for providing access to scientific journals and research articles.

\section*{Declarations}

\begin{itemize}

\item \textbf{Competing interests}\\
The authors declare that they have no competing interests.

\item \textbf{Data availability}\\
The datasets generated and/or analyzed during the current study are available on reasonable request.

\item \textbf{Code availability}\\
The code used for the analysis and reconstruction is available on reasonable request.

\end{itemize}

\printcredits

\bibliographystyle{spphys}
\bibliography{cas-refs}

\end{document}